\documentclass{elsart3p} 
\usepackage{graphicx}
\journal{Physics Letters B} 
\begin{document}
\begin{frontmatter} 
\title{Second RPA and correlated realistic interactions\thanksref{dfg} 
} 
\thanks[dfg]{Work supported by the  
Deutsche Forschungsgemeinschaft within SFB 634} 

\author{P.~Papakonstantinou\corauthref{cor}}  
\corauth[cor]{Corresponding author. 
{\em Email: }{\tt panagiota.papakonstantinou@physik.tu-darmstadt.de}}, \author{R.~Roth}

\address{Institut f\"ur Kernphysik, 
Technische Universit\"at Darmstadt,  
Schlossgartenstr.9, 
D-64289 
Darmstadt, Germany  
} 

   
\begin{abstract} 
We examine the response of closed-shell nuclei using 
a correlated interaction, 
derived with the Unitary Correlation Operator Method (UCOM) from the Argonne V18 potential, in second RPA (SRPA) calculations. 
The same correlated two-body interaction is used to derive the Hartree-Fock ground state and the SRPA equations.  
Our results 
show that the coupling of particle-hole states to higher-order configurations produces sizable effects compared with first-order RPA. 
A much improved description of the isovector dipole and isoscalar quadrupole resonances is obtained, thanks in part to the more fundamental treatment of the nucleon effective mass offered by SRPA. 
The present work suggests the prospect of describing giant resonance properties realistically and consistently within extended RPA theories. 
Self-consistency issues of the present SRPA method and residual three-body effects are pointed out. 



\end{abstract} 
\end{frontmatter} 

%

%


The problem of deriving global effective nucleon-nucleon interactions (NNI), 
based on realistic NNIs and appropriate for use in microscopic many-body theories of the nucleus, 
has been addressed recently in the framework of the Unitary Correlation Operator Method (UCOM)~\cite{RNH2004}. 
One starts from a realistic NNI and treats explicitly the short-range correlations it induces, 
so that a softened, phase-shift equivalent NNI is obtained. 
Such correlated realistic NNIs offer the possibility to exploit 
microscopic theories, ranging from Hartree-Fock (HF) and many-body perturbation theory to, e.g., no-core shell model, in a consistent and systematic way,  
as has been demonstrated in a series of applications~\cite{RHP2005,RPP2006,PPH2006}. 

The mean-field picture of the nucleus 
is empirically justified and, thanks to its simplicity, 
it is the most convenient starting point to the microscopic description of nuclear response throughout the nuclear chart. 
Small-amplitude oscillations of the density 
induced by an external field, in particular giant resonances (GRs), can 
be described self-consistently, within the Random Phase Approximation (RPA) (or quasiparticle RPA - QRPA) based upon the HF ground state 
(or HF Bogolioubov - HFB - to include pairing). 
One can go beyond first-order RPA and describe collisional damping and spreading of GRs within, e.g., second RPA (SRPA), 
or equivalent methods; 
higher-order effects are related to non-trivial parts of the two-body density matrix and can influence the position, strength and fine structure 
of GRs, as well as those of less collective low-lying states -- see, e.g., Refs.~\cite{SpW1991,Wam1988,DNS1990,LAC2004,CoB2001,She2004,TLS2004,LRV2007,Ter2007} 
and references therein.  

The input to such calculations is an effective nuclear Hamiltonian or a density functional. 
Phenomenological effective NNIs are fitted to sets of experimental data using mostly HF(B) calculations and selected (Q)RPA results. 
Their range of applicability is inevitably restricted to the selected observables and many-body methods%
\footnote{Let us note that, since no effective interactions have been fitted to SRPA calculations, mainly for computational reasons, 
consistency in the treatment of the ground and excited states is, more often than not, lost in practical applications beyond RPA.}. 

The UCOM offers an alternative path to such microscopic calculations.  
Applications in HF and RPA were discussed in Refs.~\cite{RPP2006,PPH2006,PRP2007}. 
In this work we will use the same UCOM interaction, derived from the Argonne V18 potential, in SRPA calculations of nuclear response. 
As we will see, SRPA and the UCOM make an interesting combination: Not only can SRPA accommodate more physics than first-order RPA, 
it also appears suitable for describing long-range correlations (LRC) which are excluded from the UCOM by construction. 

It is instructive to summarize the properties of the UCOM and the main conclusions of previous work. 
Short-range correlations, both central and tensor,  are explicitly described within the UCOM by means of a unitary correlation operator. 
This can be used to perform a similarity transformation of the bare nuclear Hamiltonian (or any other operator of interest). 
The resulting transformed, or ``correlated", Hamiltonian consists of a kinetic-energy term and an energy-independent two-body potential $V_{\mathrm{UCOM}}$. 
The UCOM potential is phase-shift equivalent to the original, bare one and  
has been shown to have good convergence properties~\cite{RHP2005,RPP2006}.  
Omitted three-body effects (correlations and interactions) are effectively taken into account to some extent by the parameterization of the correlators, 
while the task of describing LRC is assigned to the model space. 
The only parameters entering the formalism are in fact related to the optimal separation of state-independent short-range correlations 
and the minimization of three-body effects. 
They are fixed in the nucleus $^4$He~\cite{RHP2005}. 

The $V_{\mathrm{UCOM}}$ was employed in HF calculations in Ref.~\cite{RPP2006}. 
The binding energies and charge radii are underestimated in HF and  
the level spacing of the single-particle states is too large. 
Second-order perturbation theory (PT) constitutes 
a seemingly adequate extension of HF~\cite{RPP2006}. 
A very good description of nuclear binding energies 
was achieved within PT for nuclei from $^4$He to $^{208}$Pb, 
suggesting that the HF underbinding (about 4~MeV per nucleon) 
can be attributed to missing LRC. 
Charge radii are still underestimated within PT, implying that 
supplementing the two-body $V_{\mathrm{UCOM}}$ Hamiltonian with a three-body term (and readjusting it accordingly) 
to take into account missing effects may be necessary. 

The $V_{\mathrm{UCOM}}$ was subsequently employed in 
standard, HF-based, self-consistent RPA calculations 
to study nuclear GRs~\cite{PPH2006}. 
The isoscalar (IS) giant monopole resonance (GMR), 
the isovector (IV) giant dipole resonance (GDR), 
and the IS giant quadrupole resonance (GQR) were examined. 
A reasonable agreement with the 
experimental GMR centroid energies was achieved 
for various closed-shell nuclei, 
but the energies of the GDR and the GQR 
were overestimated by several MeV.

High GDR and GQR energies, 
as well as low single-particle level densities compared with experiment, 
are usually associated with a small nucleon effective mass, 
which in the case of the $V_{\mathrm{UCOM}}$ can be viewed as a result of missing residual LRC. 
Including explicit RPA correlations in the ground state was shown to produce small corrections~\cite{PRP2007}.  
Given that an extended model space is of great relevance  
when using the $V_{\mathrm{UCOM}}$, it is important to examine whether 
coupling of the particle-hole ($ph$) excitations to higher-order configurations 
($2p2h$ and beyond), starting with SRPA, can produce significant corrections.


We will use the SRPA as it was formulated in Ref.~\cite{Yan1987} in analogy to RPA.  
Excited states $|\lambda\rangle $ of energy $E_{\lambda}=\hbar\omega_{\lambda}$ with respect to the ground state $|0\rangle$  
are considered as combinations of $ph$ and $2p2h$ configurations. 
The corresponding creation operators $Q_{\lambda}^{\dagger}$, such that  
\begin{equation} 
|\lambda\rangle = Q_{\lambda}^{\dagger} |0\rangle \, , 
\,\,\, 
Q_{\lambda} |0\rangle = 0 \, , 
\end{equation} 
are then written as 
\begin{eqnarray}  
Q_{\lambda}^{\dagger} &=& 
\mbox{$\sum_{ph}$} X_{ph}^{\lambda} O^{\dagger}_{ph} 
- \mbox{$\sum_{ph}$} Y_{ph}^{\lambda} O_{ph} 
\nonumber \\ 
&&  
 + \mbox{$\sum_{p_1h_1p_2h_2}$} \mathcal{X}_{p_1h_1p_2h_2}^{\lambda} O^{\dagger}_{p_1h_1p_2h_2} 
\nonumber \\ 
&&  
 - \mbox{$\sum_{p_1h_1p_2h_2}$} \mathcal{Y}_{p_1h_1p_2h_2}^{\lambda} O_{p_1h_1p_2h_2} 
, 
\end{eqnarray} 
where $O^{\dagger}_{ph}$ 
creates a $ph$ state and 
$O^{\dagger}_{php'h'}$ 
creates a $2p2h$ state. 
We omit angular momentum coupling to keep the notation simple. 
The SRPA ground state, which formally is the vacuum of the annihilation operators $Q_{\lambda}$, 
is approximated by the HF ground state. 
The forward ($X$, $\mathcal{X}$) 
and backward ($Y$, $\mathcal{Y}$) 
amplitudes are the solutions of the SRPA equations  
in $ph\oplus 2p2h-$space 
\begin {equation} 
\left( \begin{array}{cc|cc}  
A                & \mathcal{A}_{12} & B  & 0 \\ 
\mathcal{A}_{21} & \mathcal{A}_{22} & 0  & 0 \\ \hline  
-B^{\ast}        &  0               & -A^{\ast} & -\mathcal{A}^{\ast}_{12} \\ 
   0             &  0               & -\mathcal{A}_{21}^{\ast} & -\mathcal{A}^{\ast}_{22} \\ 
\end{array} 
\right) 
\left( 
\begin{array}{c} 
X^{\lambda} 
\\ 
\mathcal{X}^{\lambda}  
\\ 
\hline  
Y^{\nu} 
\\ 
\mathcal{Y}^{\lambda}  
\end{array} 
\right) 
= \hbar\omega_{\lambda}  
\left( 
\begin{array}{c} 
X^{\nu} 
\\ 
\mathcal{X}^{\lambda}  
\\ 
\hline  
Y^{\nu} 
\\ 
\mathcal{Y}^{\lambda}  
\end{array} 
\right) 
\label{Esrpa}  
, \end{equation} 
where $A$ and $B$  
are the usual RPA matrices, 
$\mathcal{A}_{12}$ describes the coupling between $ph$ and $2p2h$ states 
and 
$\mathcal{A}_{22}$ contains the $2p2h$ states and their interactions. 
If we neglect the coupling amongst those states, $\mathcal{A}_{22}$ is diagonal and its elements are determined by the 
unperturbed $2p2h$ energies, 
\begin{eqnarray} 
\lefteqn{[\mathcal{A}_{22}]_{p_1h_1p_2h_2,p_1'h_1'p_2'h_2'} = } 
\nonumber \\ 
&& 
 \delta_{p_1p_1'}\delta_{h_1h_1'}\delta_{p_1p_1'}\delta_{h_1h_1'}
(e_{p_1}+e_{p_2}-e_{h_1}-e_{h_2}) 
\label{Ea22} 
,  
\end{eqnarray} 
where $e_i$ are the HF single-particle energies. 
It is interesting to note that the SRPA problem of Eq.~(\ref{Esrpa}) can be reduced to an energy-dependent eigenvalue problem of the dimension of the RPA matrix (see, e.g., Ref.~\cite{Wam1988}). Therefore, it can be viewed as an RPA problem with an energy-dependent interaction. 

The dimension $N$ of the SRPA matrix, Eq.~(\ref{Esrpa}), can be rather large. 
For the purposes of the present work we solve problems 
with $N$ up to $10^6$, but larger spaces are to be expected for heavier nuclei and larger bases. 
Fortunately, the SRPA matrix is also sparse,  
especially, but not only, when the approximation (\ref{Ea22}) is employed. 
Thus it is possible to store all its non-zero matrix elements and then use a Lanczos procedure to obtain only the 
spectrum section of interest, i.e., a couple of hundred eigenstates and eigenvalues at the lower end of the spectrum. 

By setting the coupling matrices $\mathcal{A}_{12}$ and $\mathcal{A}_{22}$ and the $2p2h$ amplitudes $\mathcal{X}$, $\mathcal{Y}$ equal to zero  
in Eq.~(\ref{Esrpa}), the usual RPA problem is retrieved. If in addition we neglect the $ph$ residual interaction (i.e., $B_{ph,p'h'}=0$ and 
$A_{ph,p'h'}=(e_p-e_h)\delta_{pp'}\delta_{hh'}$), a trivial, unperturbed problem is obtained, 
where the eigenstates $|\lambda\rangle$ are simply the $ph$ configurations at the HF level and the $Y$ amplitudes vanish. 
In all cases we may define the strength distribution of a $|ph^{-1}\rangle$ configuration as the quantity 
\begin{equation} 
S_{ph}(E_{\lambda}) = |X_{ph}^{\lambda}|^2 - |Y_{ph}^{\lambda}|^2 
.
\label{Esph}
\end{equation}  
Note that in the unperturbed case the centroid of $S_{ph}(E)$ is identical to the $ph$ energy $e_p-e_h$ and its width is zero.    

As usual, the quantities of interest are transition strength distributions $R_F(E)$ of single-particle operators ${F}^{\dagger}=\sum_{ij}f_{ij}a^{\dagger}_ia_j$,  
\begin{eqnarray} 
R_{F}(E) &=&  \sum_{\lambda} |\langle \lambda | {F}^{\dagger} |0 \rangle |^2 \delta (E-E_{\lambda})  \\ 
                     &\equiv& \sum_{\lambda} B_F(E_{\lambda}) \delta (E-E_{\lambda}) 
,
\end{eqnarray} 
and their energy moments 
\begin{equation} 
m_k=\sum_{\lambda}E_{\lambda}^k B_F(E_{\lambda})  , 
\end{equation} 
determined by the amplitudes $X$ and $Y$ through 
\begin{equation} 
\langle \lambda | F^{\dagger} |0 \rangle = \sum_{ph} [ f_{ph}{X_{ph}^{\lambda}}^{\ast} + f_{hp}{Y_{ph}^{\lambda}}^{\ast}] 
.
\end{equation} 
We will consider IS and IV transitions of definite spin and parity $J^{\pi}$,  
described by standard single-particle transition operators~\cite{PPH2006}. 

The total strenght $m_0$ and the first moment of the strength distribution $m_1$ are the same in SRPA as in RPA~\cite{AdL1988}. 
However, when based on the HF ground state, the SRPA is not fully self-consistent and symmetry-conserving, contrary to the HF-based RPA,  
as it misses a class of second-order effects related to ground-state correlations~\cite{TSA1988,ToS2004}. 
The missing effects may be important, especially for the less collective low-lying states.  
In principle, it is possible to combine the SRPA with a correlated ground state~\cite{DNS1990,TSA1988,GGC2006}, 
or employ a self-consistent Green's function method~\cite{BaD2003}, 
for a more complete theoretical treatment of nuclear excitations, but that is beyond the purposes of the present work.

\begin{figure}[t]\centering
\includegraphics[angle=-90]{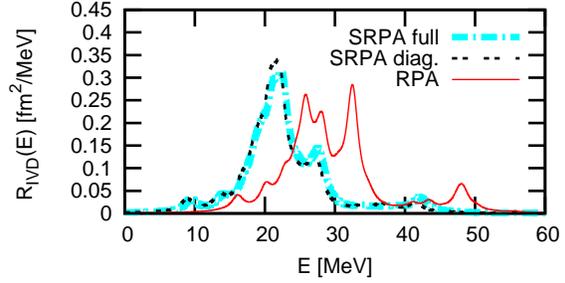} 
\caption{%
IV dipole response of $^{16}$O in a single-particle basis with $(2n+\ell)_{\max}=12$ and $\ell_{\max}=8$. Pale (cyan) dash-dotted lines: full SRPA solution; dashed (black) lines: using approximation (\ref{Ea22}); full (red) lines: RPA. 
\label{Ffull}}
\end{figure}
\begin{figure*}[t]\centering
\includegraphics[angle=-90,width=140mm]{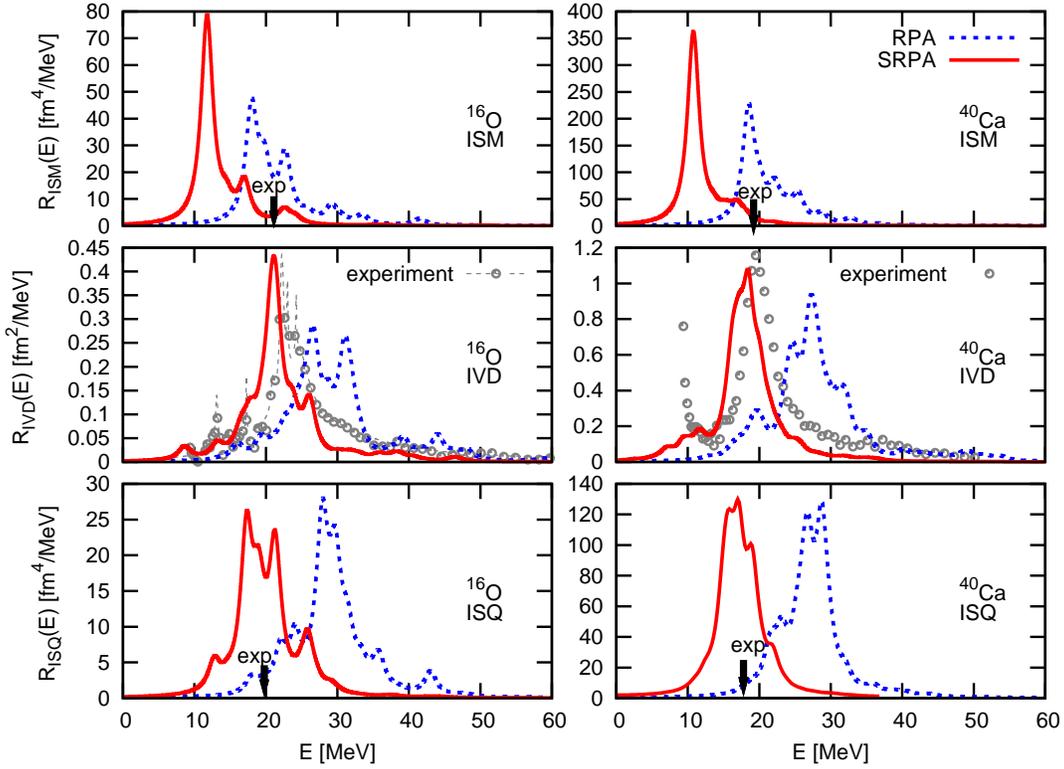} 
\caption{The IS monopole (top), IV dipole (middle) and IS quadrupole (bottom) strength distributions for the nuclei $^{16}$O (left) and $^{40}$Ca (right) within RPA (blue dashed lines) and SRPA (red full lines), compared with experiment (points, arrows). 
The calculated distributions have been folded with a Lorenzian with a width of 2~MeV. 
The experimental centroids $m_1/m_0$ of the ISM and the ISQ GRs were taken from 
Refs. \cite{LCY2001} (\nuc{16}{O}) and \cite{YLC2001} (\nuc{40}{Ca}). 
Photoabsorption cross sections were found in 
Refs. \cite{Ahr1975,LNH1987} (\nuc{16}{O}) and \cite{Vey1974} (\nuc{40}{Ca})%
(data available in \cite{CDFE}) 
and experimental IVD strength distributions 
were evaluated from those. 
\label{Fall}}
\end{figure*}

We have used the same $V_{\mathrm{UCOM}}$ as in Refs.~\cite{RHP2005,RPP2006,PPH2006} and a harmonic-oscillator single-particle basis (length parameter $b=1.7$fm) and we have examined the IS monopole (ISM), IV dipole (IVD) and IS quadrupole (ISQ) response mainly of the nuclei $^{16}$O and $^{40}$Ca in SRPA. 
The diagonal approximation, Eq.~(\ref{Ea22}) is used. It has been verified, though, that inclusion of the $2p2h$ couplings does not introduce significant corrections. An example is shown in Fig.~\ref{Ffull}.  
Note that those couplings constitute higher-order effects and their smallness suggests that corrections beyond second order are not large.

In what follows, single-particle states with radial quantum number up to $n_{\max}=6$ and orbital angular momentum up to $\ell_{\max} =6$ have been included. 
The convergence of the GR sum rules $m_0$ and $m_1$ and centroids is rather good for the present basis (within about 1-2~MeV for the centroids, being worst for the IS GMR). 

The spurious state related to the CM motion will generally not be exactly seperated from the physical spectrum, when SRPA is based on the HF ground state~\cite{ToS2004}. In order to quantify this problem, we have examined the IS dipole response. We found that spurious states appear at about 5-8~MeV. Using a transition operator of the usual radial form ($\propto r^3 - \frac{5}{3}\langle r^2 \rangle r$) and its uncorrected form ($\propto r^3$), we found that the spectrum in the GR region is not strongly affected by the choice of operator and can be considered uncontaminated. 
Further technical and numerical details regarding our SRPA implementation and consistency tests are reserved for a more extended future publication.  

\begin{figure}[t]\centering
\includegraphics[angle=-90]{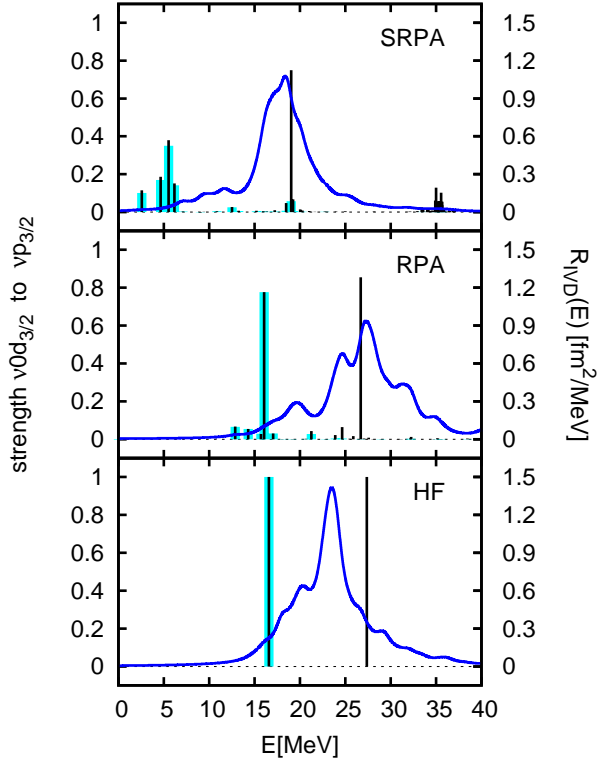} 
\caption{Fragmentation and shift of $ph$ states. Bars show how the spectroscopic strength $S_{ph}(E)$, Eq.~(\ref{Esph}), of the $ph$ configurations $|(\nu p_{3/2})(\nu 0d_{3/2})^{-1};1^{-}\rangle$ (contributing to the dipole strength) is distributed in $^{40}$Ca, within HF (bottom), RPA (middle) and SRPA (top). Thicker, pale (cyan) bars denote the distribution of $|(\nu 1p_{3/2})(\nu 0d_{3/2})^{-1};1^-\rangle$ (one shell) only. Continuous (blue) lines, corresponding to the $y-$axis scale on the right: IVD strength distribution, shown for reference. 
\label{Ffragmph}}
\end{figure}

In Fig.~\ref{Fall} we show the ISM, IVD and ISQ strength distributions for $^{16}$O and $^{40}$Ca.  
The calculated spectra (RPA and SRPA) have been folded with a Lorenzian with a width of 2~MeV, for presentation purposes. 
In all cases, we observe that the SRPA centroid energies are much lower than the RPA ones. 
The reason for the large difference between the RPA and SRPA results -- even for such collective $ph$ excitations like the GRs considered here -- is, to a large extent, the coupling of $ph$ states with virtual phonons, implicitly taken into account in SRPA.  
The inclusion of second-order configurations within SRPA effectively dresses the underlying HF single-particle states with self-energy insertions~\cite{SpW1991,Wam1988,LAC2004} and brings them closer to each other energetically, thereby lowering the $ph$ energies. 
As an illustration of the effect, we show in Fig.~\ref{Ffragmph} how the strength of the dipole configurations corresponding to a neutron $0d_{3/2}$ hole and a neutron $np_{3/2}$ particle ($n=1,2,\ldots$) 
is distributed in HF, RPA, and SRPA. In HF all the strength of each $ph$ configuration, defined in Eq.~(\ref{Esph}), is concentrated in one peak, positioned at energy equal to $e_p-e_h$. In RPA the strength of each configuration appears slightly shifted and fragmented. In SRPA the shift and fragmentation are much more pronounced. 
The shift is related to the real part of the acquired self energy and the fragmentation to the imaginary part 
and  neither can be ignored when using completely ``undressed" (with respect to LRC) HF states like the ones produced by the $V_{\mathrm{UCOM}}$. 
In this scheme the HF single-particle energies (in a similar manner as HF binding energies and radii) are viewed as auxiliary model quantities, not to be directly compared with experiment%
\footnote{A related discussion in the context of the Extended Theory of Finite Fermi Systems and 
phenomenological effective NNIs can be found in Ref.~\cite{Tse2007}.}. 
Double counting of second-order effects is thus avoided. 

\begin{figure}[t]\centering
\includegraphics[angle=-90]{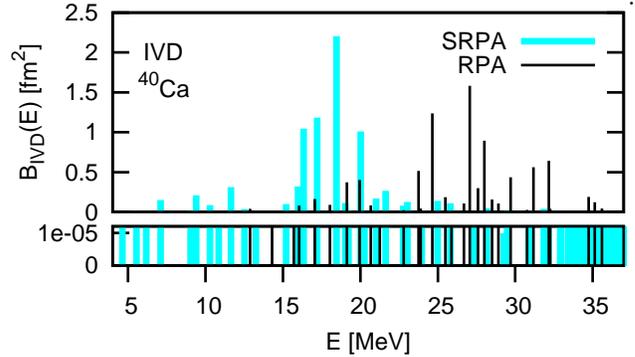}. 
\caption{Fragmentation in SRPA. Both panels show the IVD transition strength distribution of $^{40}$Ca in RPA (black bold bars) and SRPA (pale cyan bars). In the lower panel the $y-$scale is different, in order for the large amount of weaker SRPA states to become visible. 
\label{Ffragm}}
\end{figure}

In Fig.~\ref{Fall} we also observe that, in some cases, the resonance width seems to be smaller in SRPA than in RPA, contrary to what one might expect. 
The reason appears to be again an overal compression of the underlying $ph$ spectrum
that competes with the increased fragmentation. 
Fragmentation does occur in SRPA, as is illustrated in Fig.~\ref{Ffragm}. 

\begin{figure}[t]\centering
\includegraphics[angle=-90]{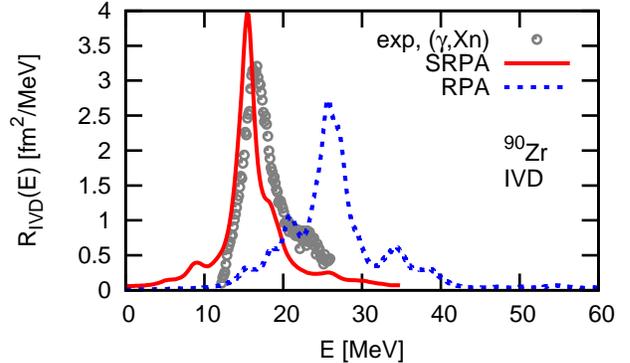} 
\caption{IVD transition strength distribution of $^{90}$Zr in RPA and SRPA compared with $(\gamma, Xn)$ data from \cite{Lep1970,CDFE}. 
\label{FZr}}
\end{figure}
Let us now discuss our results in comparison with experiment.   
In the middle panels of Fig.~\ref{Fall} our IVD strength distributions are shown along with those extracted from experimental data (there has been no ad hoc renormalization imposed). 
We observe that the IV GDR is more realistically reproduced by SRPA than by RPA. 
Its centroid energy is somewhat underestimated. 
These trends seem to persist in the heavier nucleus $^{90}$Zr, as shown in Fig.~\ref{FZr}.  
In the lower panels of Fig.~\ref{Fall} we show the ISQ strength distributions. The RPA and SRPA results are shown and the experimental centroids of the IS GQR are indicated. The SRPA results agree very well with experiment, 
suggesting that the  
coupling to higher-order configurations restores a realistic nucleon effective mass to a large extent. 
For the distribution widths, which we evaluate from the energy moments in the resonance region as 
$\sigma = \sqrt{m_2/m_0 - (m_1/m_0)^2}$, 
we find 3.8~MeV (2.5~MeV) for $^{16}$O ($^{40}$Ca), smaller than, but comparable with, the experimental widths of 5.1~MeV (2.9~MeV)~\cite{LCY2001,YLC2001}. SRPA is a suitable theory for describing the spreading width of resonances, but one should keep in mind the important role of the continuum in light and medium-heavy nuclei, which is not properly described by our method and model space. 
 
In the upper panels of Fig.~\ref{Fall} the ISM strength distributions are shown. The energies of the IS GMR are underestimated within SRPA. 
This may be another indication that there are missing three-body effects. 
Normally, residual three-body corrections should affect the IS GMR most of all, since it is a compression mode. They should affect less strongly the IV GDR, where the nuclear interior plays a lesser role, and less the IS GQR, which is a surface mode. These physical arguments could serve as a guide for the construction of an appropriate effective three-body term to supplement the two-body $V_{\mathrm{UCOM}}$. In any case, the three-body term shall depend on the bare interaction used, as well as the correlation operators applied to it~\cite{RHP2005}. Another source of problems could of course be the inherent inconsistencies of the present SRPA formalism, which will be investigated in future work.


In summary, we have used a correlated interaction, 
derived within the UCOM from the Argonne V18 potential, in SRPA calculations of nuclear response. 
Our results for the nuclei $^{16}$O and $^{40}$Ca show that the coupling to higher-order configurations produces sizable effects, compared with first-order RPA. An improved description of the IVD and ISQ resonances is obtained. Through a more fundamental treatment of the nucleon effective mass, the UCOM-based SRPA method seems to enable a simultaneous description of IVD and ISQ GRs. 
Our correlated interaction underestimates the energy of the ISM GR, though, pointing to missing three-body effects. Efforts to construct appropriate three-body potentials to complement our two-body correlated potentials are under way.  

The present work suggests the prospect of describing GR centroids and structure realistically and consistently, 
within extended RPA theories like SRPA. 
The crucial points in this context are that the $V_{\mathrm{UCOM}}$ interaction does not parameterize LRC, which are instead described by the SRPA, 
and of course the good convergence properties of the $V_{\mathrm{UCOM}}$, which should render the SRPA (or equivalent) model space flexible enough to describe residual correlations. 
More systematic calculations are planned for the immediate future, in order to assess and explore this possibility. These shall include low-lying collective states.  
Resolving the self-consistency issues of the present SRPA method is planned as well. 
%
%

%
%


\begin{thebibliography}{10}
\expandafter\ifx\csname url\endcsname\relax
  \def\url#1{\texttt{#1}}\fi
\expandafter\ifx\csname urlprefix\endcsname\relax\def\urlprefix{URL }\fi

\bibitem{RNH2004}
R.~Roth, T.~Neff, H.~Hergert, H.~Feldmeier, Nucl. Phys. A745 (2004) 3.

\bibitem{RHP2005}
R.~Roth, H.~Hergert, P.~Papakonstantinou, T.~Neff, H.~Feldmeier, Phys. Rev. C72
  (2005) 034002.

\bibitem{RPP2006}
R.~Roth, P.~Papakonstantinou, N.~Paar, H.~Hergert, T.~Neff, H.~Feldmeier, Phys.
  Rev. C73 (2006) 044312.

\bibitem{PPH2006}
N.~Paar, P.~Papakonstantinou, H.~Hergert, R.~Roth, Phys. Rev. C74 (2006)
  014318.

\bibitem{SpW1991}
J.~Speth, J.~Wambach, in: J.~Speth (Ed.), Electric and Magnetic Giant
  Resonances in Nuclei, 1991, p.~1.

\bibitem{Wam1988}
J.~Wambach, Rep. Prog. Phys. 51 (1988) 989.

\bibitem{DNS1990}
S.~Dro{$\dot{\mathrm z}$}d{$\dot{\mathrm z}$}, S.~Nishizaki, J.~Speth,
  J.~Wambach, Phys. Rep. 197 (1990) 1.

\bibitem{LAC2004}
D.~Lacroix, S.~Ayik, P.~Chomaz, Prog. Part. Nucl. Phys. 52 (2004) 497.

\bibitem{CoB2001}
G.~Col\`o, P.~Bortignon, Nucl. Phys. A687 (2001) 282c.

\bibitem{She2004}
A.~Shevchenko, et~al., Phys. Rev. Lett. 93 (2004) 122501.

\bibitem{TLS2004}
N.~Tsoneva, H.~Lenske, C.~Stoyanov, Nucl. Phys. A731 (2004) 273.

\bibitem{LRV2007}
E.~Litvinova, P.~Ring, D.~Vretenar, Phys. Lett. B647 (2007) 111.

\bibitem{Ter2007}
G.~Tertychny, et~al., Phys. Lett. B647 (2007) 104.

\bibitem{PRP2007}
P.~Papakonstantinou, R.~Roth, N.~Paar, Phys. Rev. C75 (2007) 014310.

\bibitem{Yan1987}
C.~Yannouleas, Phys. Rev. C35 (1987) 1159.

\bibitem{AdL1988}
S.~Adachi, E.~Lipparini, Nucl. Phys. A489 (1988) 445.

\bibitem{TSA1988}
K.~Takayanagi, K.~Shimizu, A.~Arima, Nucl. Phys. A477 (1988) 205.

\bibitem{ToS2004}
M.~Tohyama, P.~Schuck, Eur. Phys. J. A19 (2004) 203.

\bibitem{GGC2006}
D.~Gambacurta, M.~Grasso, F.~Catara, M.~Sambataro, Phys. Rev. C73 (2006)
  024319.

\bibitem{BaD2003}
C.~Barbieri, W.~Dickhoff, Phys. Rev. C 68 (2003) 014311.

\bibitem{LCY2001}
Y.-W. Lui, H.~Clark, D.~Youngblood, Phys. Rev. C64 (2001) 064308.

\bibitem{YLC2001}
D.~Youngblood, Y.-W. Lui, H.~Clark, Phys. Rev. C63 (2001) 067301.

\bibitem{Ahr1975}
J.~Ahrens, H.~Borchert, K.~Czock, H.~Eppler, H.~Gimm, H.~Gundrum, M.~Kroning,
  P.~Riehn, G.~Sita~Ram, A.~Zieger, B.~Ziegler, Nucl. Phys. A251 (1975) 479.

\bibitem{LNH1987}
S.~LeBrun, A.~Nathan, S.~Hoblit, Phys. Rev. C35 (1987) 2005.

\bibitem{Vey1974}
A.~Veyssi\`ere, H.~Beil, R.~Berg\`ere, P.~Carlos, A.~Lepr\^etre, A.~de~Miniac,
  Nucl. Phys. A227 (1974) 513.

\bibitem{CDFE}
CDFE database, \\ http://cdfe.sinp.msu.ru/services/gdrsearch.html.

\bibitem{Tse2007}
V.~Tselyaev, et~al., Phys. Rev. C 75 (2007) 014315.

\bibitem{Lep1970}
A.~Lepr\^etre, H.~Beil, R.~Berg\`ere, P.~Carlos, A.~Veyssi\`ere, M.~Sugawara,
  Nucl. Phys. A175 (1971) 609.

\end{thebibliography}
\end{document}